\documentclass{aastex631}

\begin{document}

\newcommand{\vdag}{(v)^\dagger}
\newcommand{\rno}{RNO~54}

\received{September 11, 2023}
\revised{October 25, 2023}




\title{RNO 54: A Previously Unappreciated FU Ori Star}

\author{Lynne A. Hillenbrand} 
\affiliation{Department of Astronomy, MC 249-17, California Institute of Technology, Pasadena, CA 91125, USA}
\email{lah@astro.caltech.edu}

\author{Adolfo Carvalho}
\affiliation{Department of Astronomy, MC 249-17, California Institute of Technology, Pasadena, CA 91125, USA}

\author{Jan van Roestel} 
\affiliation{Department of Astronomy, MC 249-17, California Institute of Technology, Pasadena, CA 91125, USA}
\affiliation{Current Address: Anton Pannekoek Institute for Astronomy, University of Amsterdam, NL-1090 GE Amsterdam, Netherlands}

\author{Kishalay De} 
\affiliation{Department of Astronomy, MC 249-17, California Institute of Technology, Pasadena, CA 91125, USA}
\affiliation{Current Address: Kavli Institute for Astrophysics and Space Research, Massachusetts Institute of Technology, Cambridge, MA, USA)}

\begin{abstract}
We present evidence in support of the hypothesis that 
the young stellar object RNO 54 is a mature-stage FU Ori type source. 
The star was first cataloged as a ``red nebulous object" in the 1980s
but appears to have undergone its outburst prior to the 1890s.  
Present-day optical and near-infrared spectra 
are consistent with those of other FU Ori type stars, both in the details
of spectral line presence and shape, and in the overall change in spectral type 
from an FGK-type in the optical, to the M-type presented in the near-infrared.  
In addition, the spectral energy distribution of \rno\ is well-fit 
by a pure-accretion disk model with parameters:
$\dot{M} = 10^{-3.45\pm0.06}$ $M_\odot$ yr$^{-1}$, $M_* = 0.23\pm0.06 \ M_\odot$, and $R_\mathrm{inner} = 3.68\pm0.76 \ R_\odot$, 
though we believe $R_\mathrm{inner}$ is likely close to its upper range of $4.5 R_\odot$
in order to produce a $T_\mathrm{max} = 7000$ K that is consistent with the optical to near-infrared spectra. 
The resulting $L_\mathrm{acc}$ is $\sim 265 \ L_\odot$.
To find these values, we adopted a source distance $d=1400$ pc and
extinction $A_V=3.9$ mag, along with disk inclination $i=50$ deg 
based on consideration of confidence intervals from our initial disk model, 
and in agreement with observational constraints. 
The new appreciation of a well-known source as an FU Ori type object
suggests that other such examples may be lurking in extant samples.
\end{abstract}


\section{Introduction}

FU Ori stars are a rare breed of young stellar object (YSO); 
see \cite{herbig1977,hartmann1996} and \cite{fischer2022} for reviews.
Like other YSOs, FU Ori stars have significant luminosity in the infrared to millimeter 
portion of their spectral energy distributions, and they can be 
closely associated with optical and/or near-infrared nebulae.  
Such observational features are generally consistent with a young star 
that is surrounded by a substantial amount of circumstellar dust, 
distributed over of tens to thousands of astronomical units,
that often heavily obscures the central proto- or pre-main sequence star.

However, beyond the standard YSO features, 
members of the FU Ori class have certain additional observational characteristics
that distinguish them.  Specifically, FU Ori optical spectral types 
are those of FGK-type giants to supergiants, while near-infrared spectra 
exhibit the features of M-type giants to supergiants. The
distinctive wavelength-dependent temperature pattern, as well as the broad and ``flat-bottomed"
absorption line profiles in FU Ori photospheres are such that explanations such as starspots 
or a multiple star system, can be discarded.
In addition, all FU Ori type stars show strong wind/outflow signatures
as evidenced by asymmetrically blue-shifted absorption or P Cygni profiles in certain lines.

The interpretion that FU Ori objects are YSOs in a state of prolonged accretion outburst
is commonly accepted.  During initiation of the outburst, 
a protoplanetary disk surrounding a young star increases 
its rate of mass transfer from the disk onto the central star,
transitioning from low-state accretion within the broad range of  
$\sim 10^{-8}$ to $10^{-11}\ M_\odot$ yr$^{-1}$, to enhanced levels 
in the high-state approaching $10^{-4}$ to $10^{-5}\  M_\odot$ yr$^{-1}$. 
FU Ori spectral energy distributions (SEDs) and even more convincingly, 
their high-dispersion spectra
\citep{kenyon1988,welty1992,rodriguez2022,carvalho2023a,liu2022}, 
can both be reproduced by pure-accretion disk models 
having a temperature and a velocity gradient with radius.

The prototype, FU Ori, and several other members of the FU Ori class 
(e.g. HBC 722) are located in or near well-known star forming regions. 
Other examples (e.g. Gaia 17bpi) are more isolated 
and associated with small dark clouds but not significant star forming regions. 

In this paper we present evidence regarding a relatively isolated YSO that was documented as such
in the literature decades ago, but appears to have been overlooked all of these years
as a bona fide FU Ori type star.
Although no accretion outburst was observed in \rno, 
the object is empirically similar to classical members of the class,
and both its spectral energy distribution and its high-dispersion spectrum
can be well-fit by an accretion disk model.
\rno\ is therefore a neglected, but likely FU Ori star.
This same conclusion was independently reached in a recent paper by \cite{magakian2023}.

\section{Background on the Young Stellar Object RNO 54}

\rno, also known as PDS 120, IRAS 05393+2235 and GN 05.39.2,
is located at 05:42:21.24 +22:36:47.1 (J2000.). 
The parallax, proper motions, and radial velocity of \rno\ 
seem to make it kinematically isolated from nearby stars on the sky.
No kinematically similar sources could be found in Gaia DR3 data \citep{gaiadr3}. Furthermore,
no young stellar objects could be identified in nearby projection in SIMBAD \citep{wenger2000}.
\rno\ thus appears to be a relatively isolated YSO.

\rno\ has a flat-spectrum type Class I spectral energy distribution. 
It also has a cometary or ring-like nebula similar to those associated with the YSOs 
V1515 Cyg, FU Ori, and Parsamian 21, among others, 
but differently shaped than the conical nebulae exhibited by e.g. PV Cep, R Mon and V1647 Ori.  
Although its sky location is just north of the greater Orion large star-forming complex, towards the galactic plane,
\rno\ has a Gaia parallax 
that corresponds to a much further (3.5 times) distance of $1.40\pm 0.05$ kpc.
While the error bar is only a few percent, the somewhat high RUWE value of 1.4 
could make the kinematic parameters untrustworthy, but is also explainable by the nebulous nature of the source.  
Other Gaia-derived parameters include log $T_{eff} = 6700$ K and log g = 2.32,
and $A_o$ (similar to $A_V$) = 4.21 mag.  Besides the temperature and gravity 
expected for an FU Ori type object, \rno\ also has a high luminosity reported by Gaia.
The $LUM-FLAME$ value is 780 $L_\odot$, which is derived from
the observed Gaia G magnitude, the implied $A_G$ extinction, and a bolometric correction
appropriate to the spectral type.  

In the literature, \rno\ was first noted by \cite{cohen1980} as a Red Nebulous Object 
with cometary morphology, that was bright in the near-infrared and had H$\alpha$ emission.
\cite{cohen1980} assigned a spectral type of F5 II and an extinction $A_V = 3.8\pm 0.3$ mag.
\cite{goodrich1987} suggested that nebulae, which had already been associated with FU Ori objects by \cite{herbig1977},
could be used as signposts to identify additional sources that might be ``either post- or pre-outburst FU Orionis stars."
Indeed, \cite{goodrich1987} specifically mentioned RNO 54 as an example, 
and also provided spectra typed as early G Ib or II.
Later, \cite{torres1995} included RNO 54 in a table of ``probable post-FU Ori stars",
calling it an F8 II and noting its $H\alpha$ and \ion{Li}{1} 6707 \AA\ profiles.

No actual outburst has been detected for \rno, but if indeed an FU Ori star,
the outburst is constrained to have occurred prior to its detection on the epoch 1951.85 
Palomar sky survey plate used by \cite{cohen1980}.  The outburst was also likely prior to the 
Astrographic Catalog, for which the position is reported by \cite{fresneau1983} 
at epoch 1893.08 with photographic (blue) magnitude 13.0, close to the current value.

Since these early papers, \rno\ has been included only in survey studies, 
namely the \cite{magakian2003} catalog of reflection nebulae,
the \cite{lumsden2013} Red MSX Source Survey, 
an OH maser catalog by \cite{engels2015}, 
and the \cite{lopez2021} survey of large-scale Herbig Haro flows
associated with IRAS-identified young stellar objects.
As this manuscript was being prepared, a single-object study by \cite{magakian2023} appeared.
These authors come to the same conclusion we do about RNO 54: that it is very likely an FU Ori type star. 
Both papers reinforce the original speculation of \cite{goodrich1987}.

\section{New Spectral Data and Comparison to Benchmark FU Ori Stars}

Based on a re-read of the \cite{goodrich1987} and \cite{cohen1980} papers 
undertaken as part of general scholarship, we had suspicions about the nature 
of \rno, and sought to investigate whether modern data could confirm an FU Ori interpretation.
During fall of 2020, we thus obtained optical and near-infrared spectra 
in order to investigate this hypothesis in detail.

\subsection{Data Acquisition and Reduction}

Our optical spectrum was obtained using the W.M. Keck Observatory and Echellette Spectrograph and Imager \citep[ESI][]{sheinis2002}
by J. van Roestel on 2020-09-12 UT in 1.2\arcsec\ seeing. 
The 1.0\arcsec\ slit was used with 2-pixel spatial binning, 
yielding an effective spectral resolution of $R\approx 7,500$ 
in a single 300 sec integration.
Data reduction was kindly performed by T. Kupfer using the MAKEE package\footnote{} written by T. Barlow.
The one-dimensional spectra cover the range $\approx 3,900-10,000$ \AA\ with some overlap between orders.


Our infrared spectrum was obtained using the NASA's IRTF and SpeX instrument \cite{rayner2003} by K. De on 2020-09-22 UT.
The 0.5\arcsec\ slit was used yielding an effective spectral resolution of $R\approx 1,200$ in the SXD mode, covering $\sim 0.7-2.5~\mu$m.
Dithered exposures were used to obtain a total integration time of 180 sec on source.
The spectral images were reduced using the $Spextool$ package \cite{cushing2004}
with flux calibration performed using the $XTellCor$ package \cite{vacca2003}.


\subsection{Basic Spectral Analysis}

\begin{figure}[!t]
\includegraphics[width=0.5\linewidth, trim={0 0 0.85cm 0},clip]{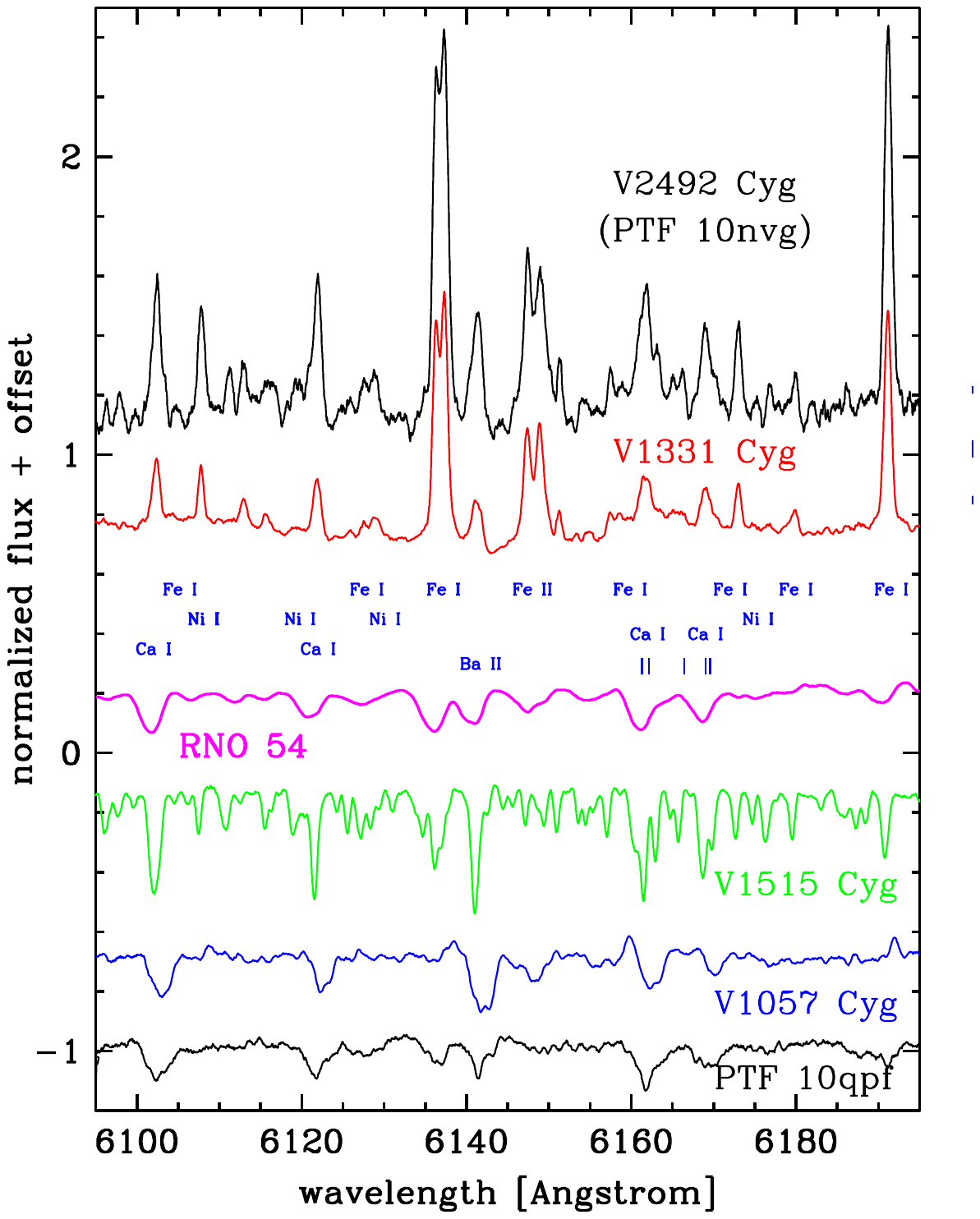}
\caption{
Optical spectrum of \rno\ (magenta) compared to the FU Ori objects 
HBC 722 (PTF 10qpf), V1057 Cyg, and V1515 Cyg,  
and to the continuum-plus-emission objects V1331 Cyg and V2492 Cyg (PTF 10nvg). 
\rno\ has an absorption spectrum similar to
the broad-line FU Ori stars V1057 Cyg and HBC 722 (PTF 10qpf);
V1515 Cyg has narrower and deeper lines than the others, due to its nearly face-on inclination.
\rno\ thus appears to be an extreme accretor, and does not have the emission line spectrum
of mere atypically-high accretion rate sources.
}
\label{fig:optspec}
\end{figure}

\begin{figure}[!t]
\includegraphics[width=0.95\linewidth,trim={0.85cm 0 0 0},clip]{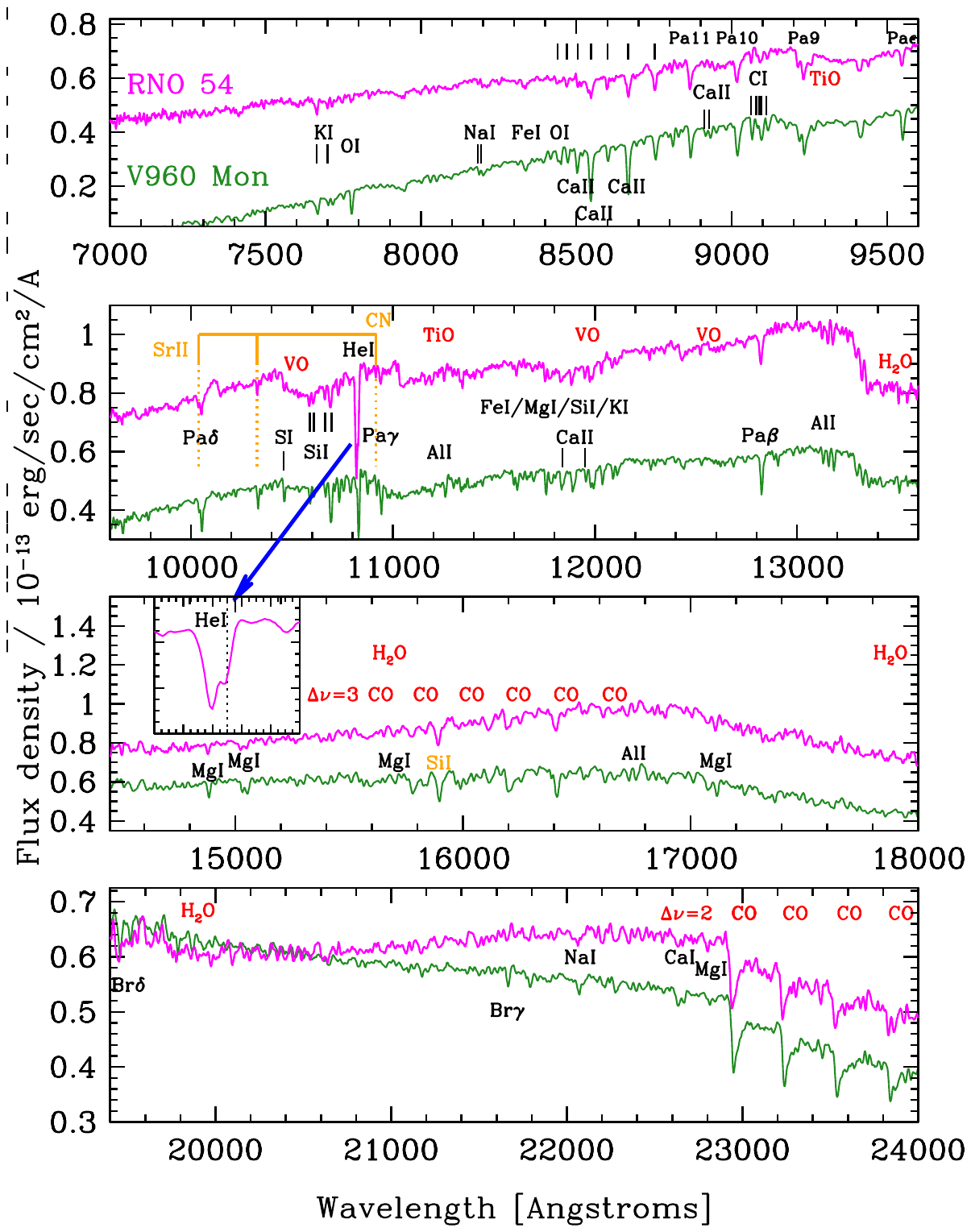}
\caption{
Infrared spectrum of \rno\ compared to that of V960 Mon. 
The two sources do not have similar continuum shapes, 
with V960 Mon artifically adjusted to attempt to match the reddening of \rno.
However, both sources have absorption features
from a mix of molecules (TiO, VO, H$_2$O, CO; red), low-excitation (\ion{Mg}{1}, \ion{Ca}{1},  \ion{Al}{1}), 
and higher-excitation (\ion{Si}{1}, \ion{S}{1}, \ion{C}{1}, \ion{Ca}{2}) neutral and ionized species (black), 
as well as the \ion{Sr}{2}, CN, and \ion{Si}{1} indicators of low surface gravity (orange).  
In addition, both sources show broad blueshifted absorption in the \ion{He}{1} 10830 line (see inset for \rno),
indicating a strong wind.
}
\label{fig:irspec}
\end{figure}

A segment of the optical spectrum appears in Figure~\ref{fig:optspec}
in comparison to other FU Ori type stars and young stellar objects with continuum-plus-emission spectra.
The full infrared spectrum is shown in Figure~\ref{fig:irspec} in comparison to the bona fide FU Ori star V960 Mon. 

Our optical and near-infrared spectra of \rno\ enable investigation of
properties of the ``photosphere", which appears to be dominated by absorption lines
produced in an accretion disk. We can also study the spectral signatures of outflowing material in winds.  

The Keck/ESI and the IRTF/SpeX spectra were taken ten days apart, 
and are essentially identical in the wavelength range of their overlap 
when convolved to the same spectral resolution. 
There is no significant variability in either absorption or emission line strengths and profiles.
Comparing our data with spectral standards and with model atmospheres 
implies a heliocentric radial velocity of about +15 km/s for \rno.  

Source extinction can also be probed spectroscopically, given that
the optical spectrum of \rno\ displays clear DIB absorption at 5850, 6270, 6379, and 6614 \AA. 
Relations established in \cite{carvalho2022} can be applied to the 6614 \AA\ DIB strength of $332 \pm 2$ m\AA,
to infer a reddening value of E(B-V) = 1.35 mag and thus a visual extinction estimate of $A_V=4.2$ mag.  
We note that this is identical to the Gaia DR3 value quoted above. 

\subsection{Signatures of an FU Ori Type Spectrum}

\begin{figure}[!t]
\includegraphics[width=0.475\linewidth]{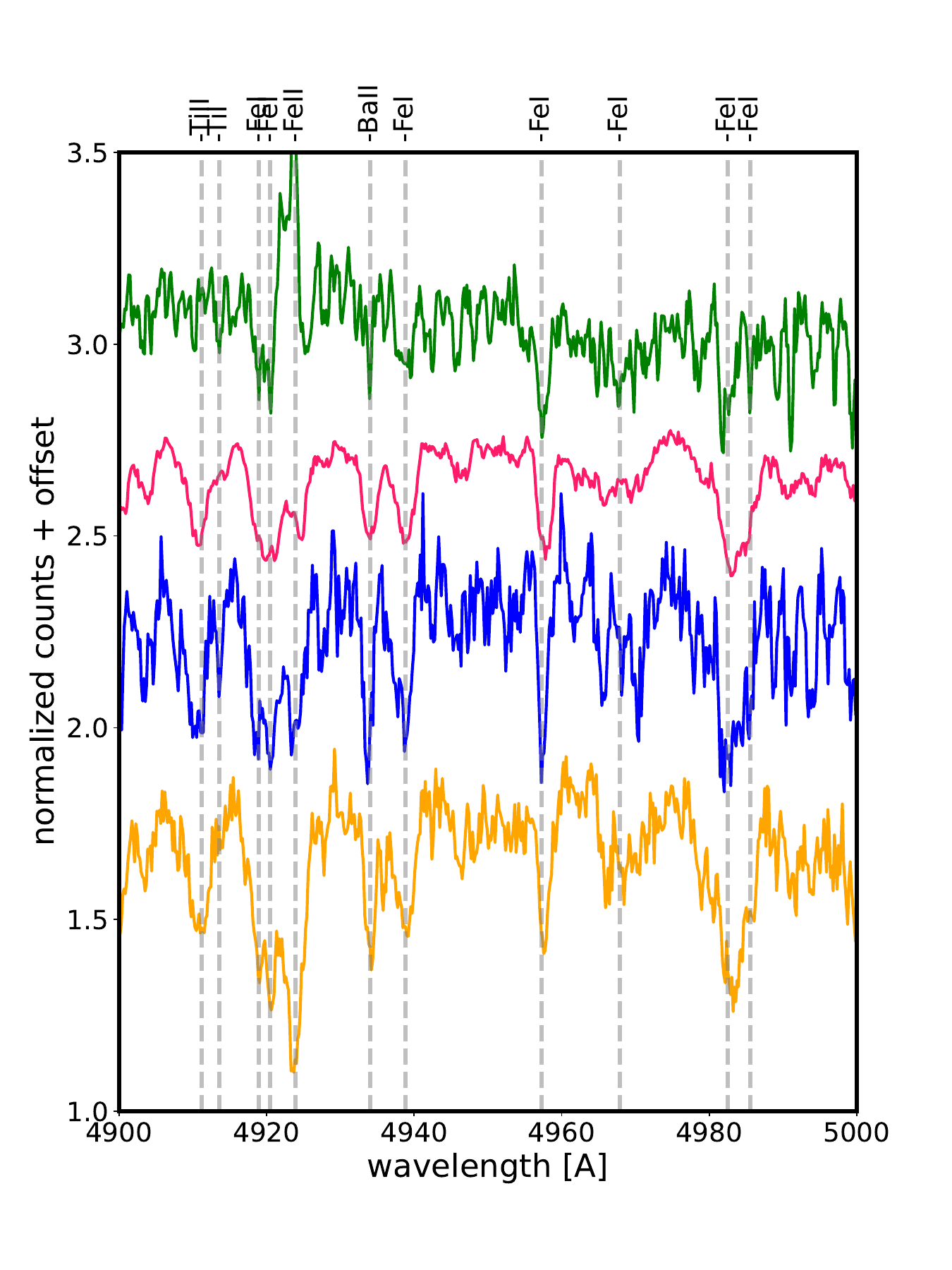}	
\includegraphics[width=0.475\linewidth]{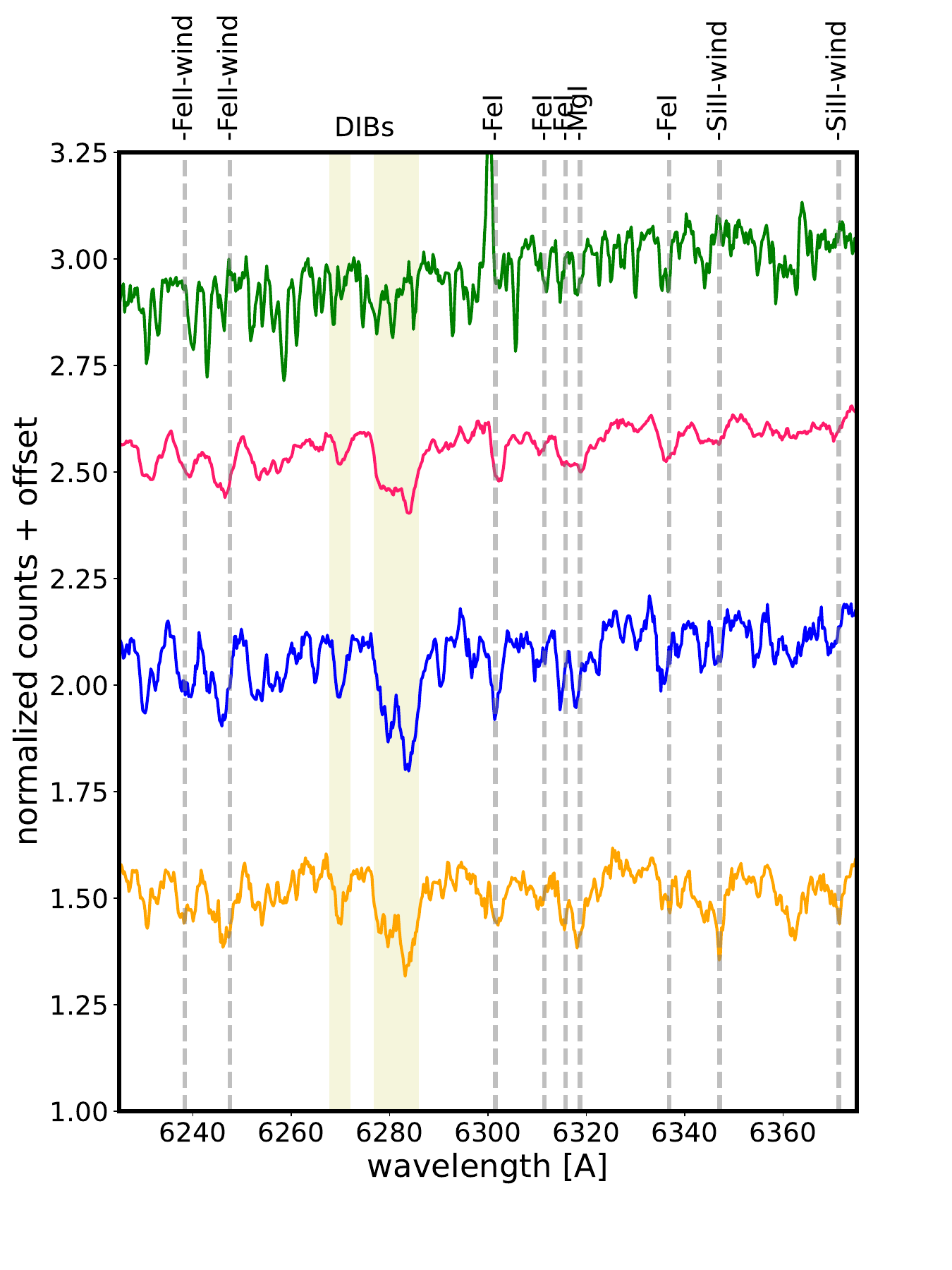}	
\includegraphics[width=0.475\linewidth]{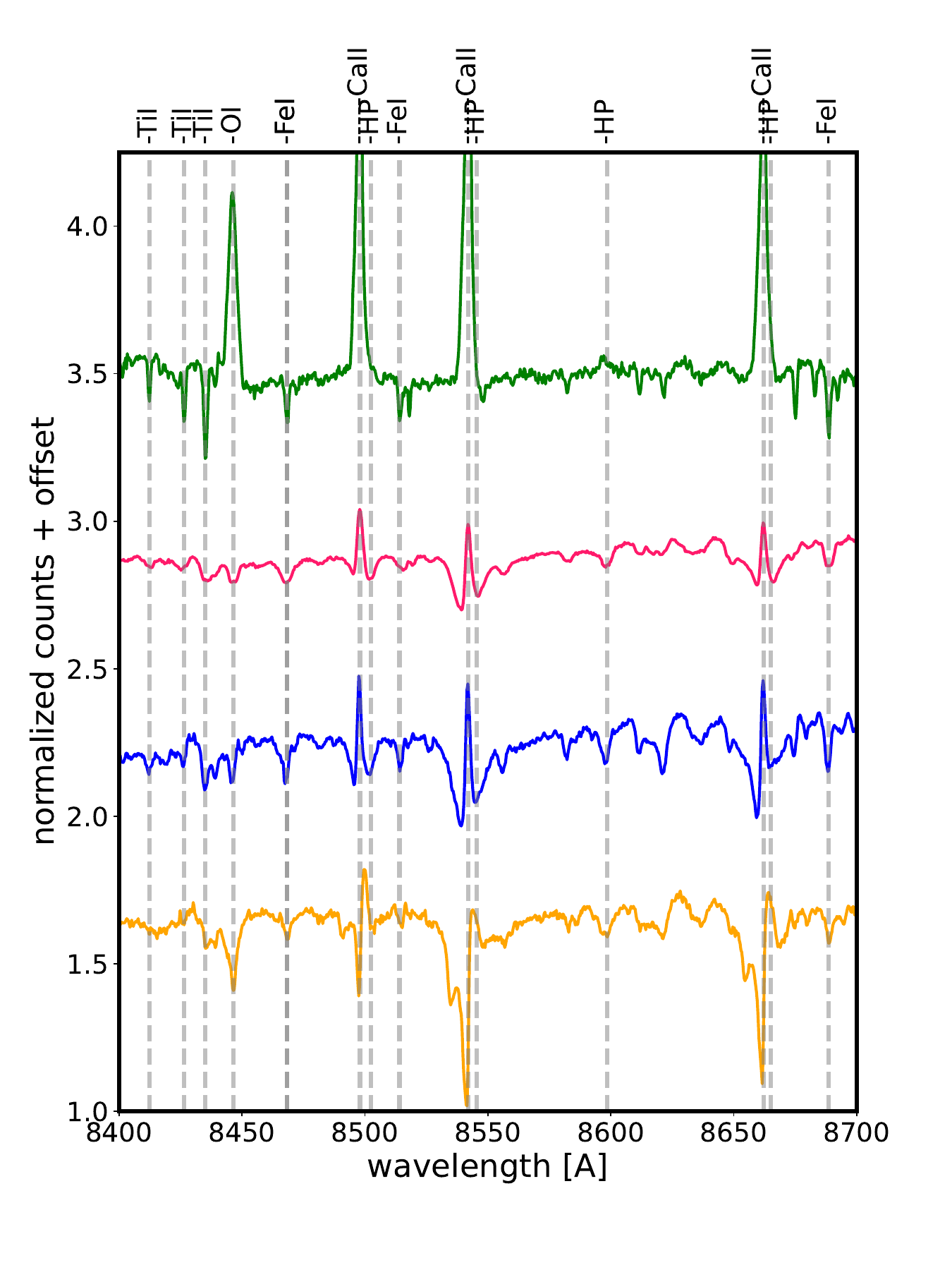}	
\includegraphics[width=0.475\linewidth]{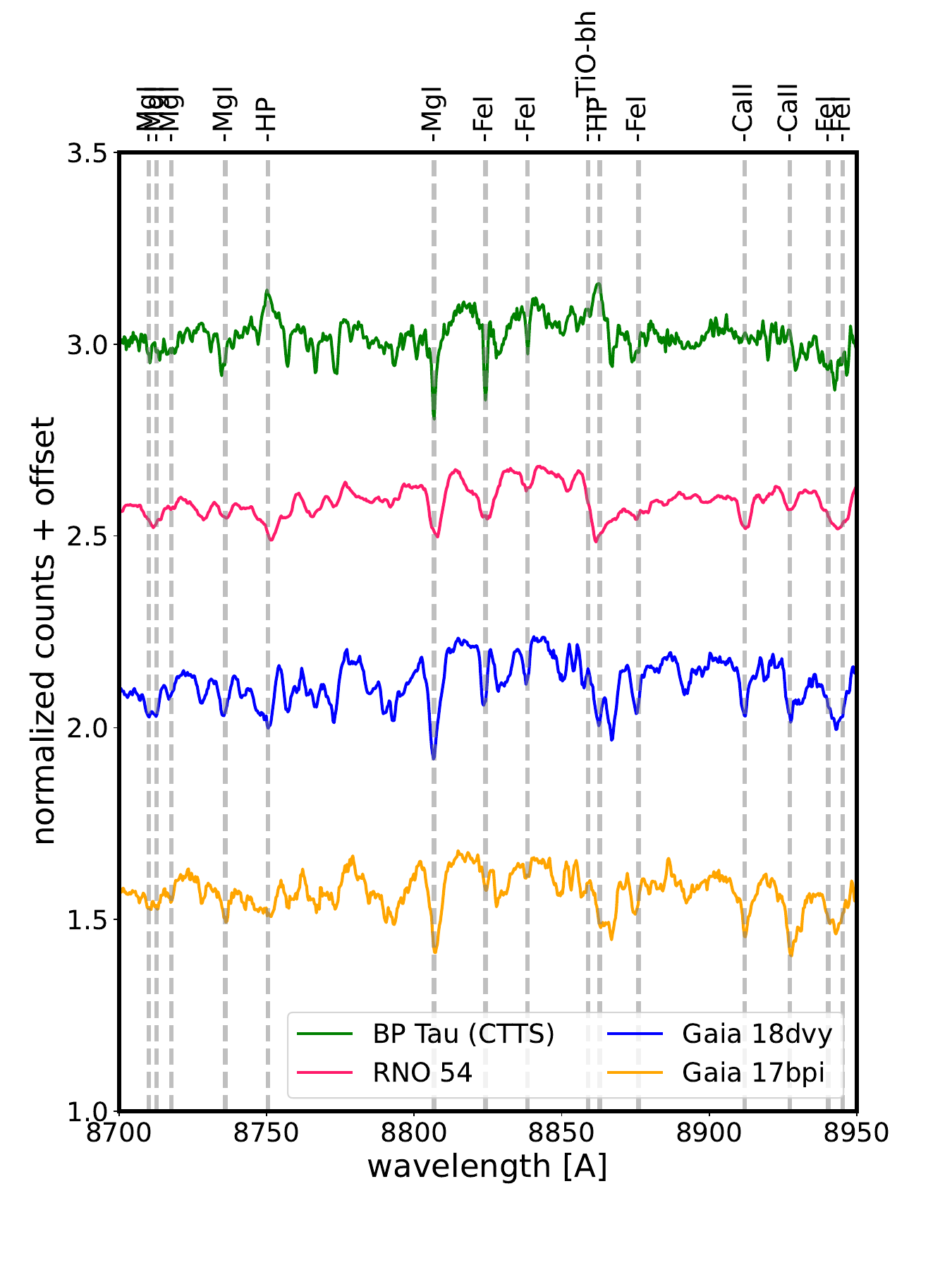}	
\caption{
Portions of the optical spectrum of \rno\ (magenta) 
compared to the recent FU Ori outburst sources 
Gaia 18dvy (blue) and Gaia 17bpi (orange), 
as well as to a non-outbursting but heavily accreting classical T Tauri star, BP Tau (green).
Spectra have been normalized and offset, for clarity.
The lower three objects are similar in terms of their line presence, depth,
and width, exhibiting absorption from a mix of low-excitation 
and higher-excitation neutral and ionized species. 
The top spectrum, by contrast, is less complicated and while similar to the 
FU Ori objects over small spectral ranges, has fewer high-excitation lines, 
somewhat deeper lines, and much narrower lines.
}
\label{fig:esi}
\end{figure}

The spectrum of \rno\ shows both strong wind and outflow signatures
evident as P Cygni type lines, 
and symmetrically broadened disk-like absorption lines. 
Both features are associated with bona fide members of the FU Ori class of YSOs.

\begin{figure}[!t]
\includegraphics[width=\linewidth]{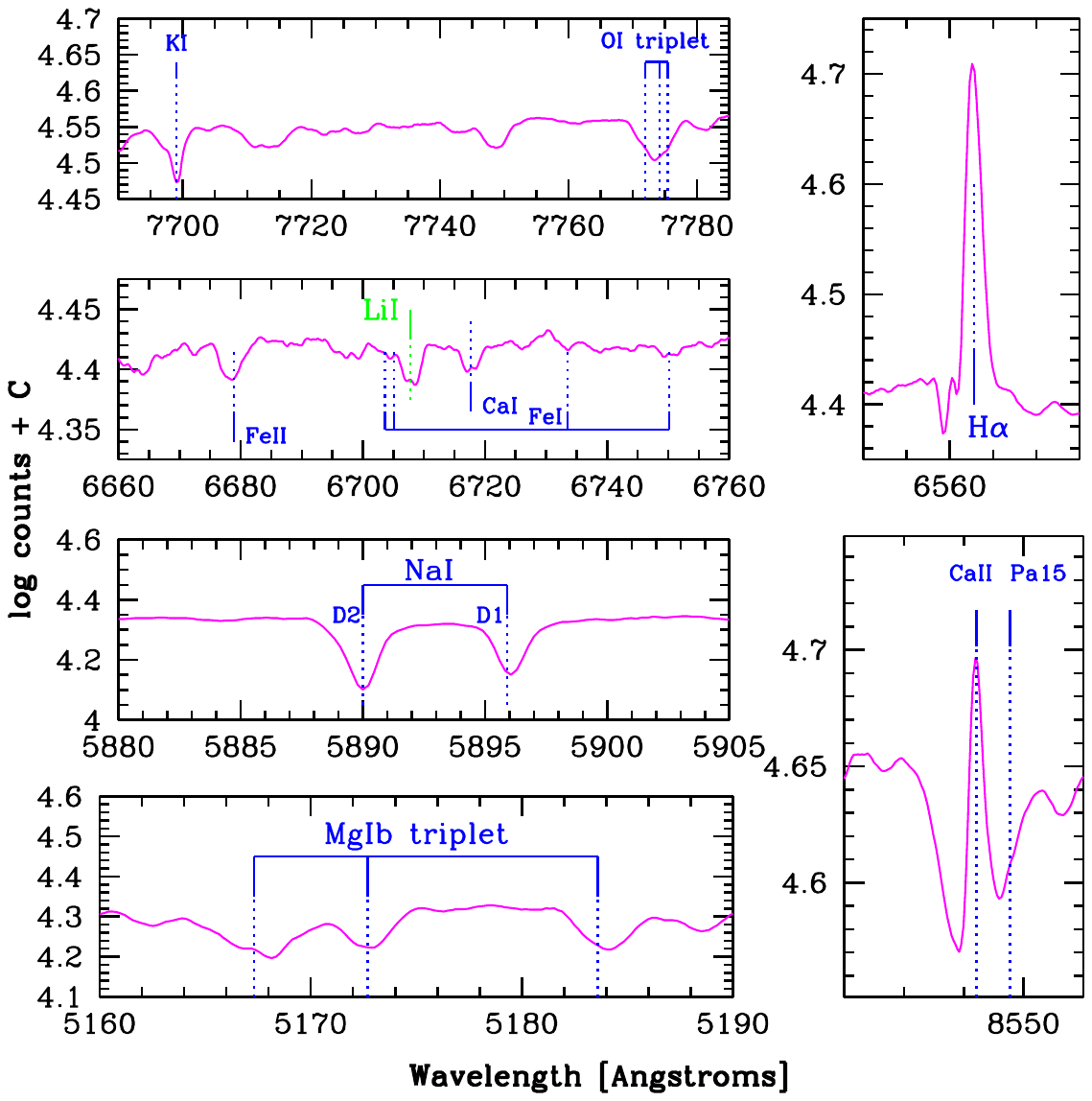}	
\caption{Segments of the optical spectrum of \rno\ emphasizing lines showing wind signatures, including 
the \ion{K}{1} 7669,7699 \AA\ doublet, the \ion{O}{1} 7773 \AA\ triplet,
the deep \ion{Na}{1} D doublet, 
the \ion{Mg}{1}b doublet, H$\alpha$, and the \ion{Ca}{2} triplet. 
The \ion{Li}{1} 6707 \AA\ signature of stellar youth is also highlighted,
showing line splitting consistent with a disk profile, similar to \ion{Ca}{1} 6717 \AA.
}
\label{fig:profs}
\end{figure}

\subsubsection{Disk Lines}

Figure~\ref{fig:esi} demonstrates that the optical spectrum of \rno\ is close 
in appearance to the recent FU Ori outburst sources
Gaia 17bpi \citep{hillenbrand2018} and Gaia 18dvy \citep{szegedi-elek2020}, 
and unlike the low-state accreting classical T Tauri star BP Tau. 
The latter has 
more low-excitation absorption lines from species like \ion{Ti}{1} and \ion{Fe}{1} visible,
whereas the three FU Ori stars (counting \rno) have low-excitation lines as well as
strong absorption from high-excitation, hotter lines such as 
\ion{Fe}{2}, seen especially towards the bluer range of spectrum (4000-5500 \AA).
BP Tau has some of these same lines in emission, e.g. \ion{Fe}{2} 4924, 5018, 5169 \AA.
In addition, BP Tau shows considerably less broadening than the FU Ori sources,
with a $v~sin~i\approx 13$ km/s that is typical for a T Tauri star.
For \rno, absorption widths are around 85 km/s (HWHD) for the disk lines.

The optical photosphere of \rno\ (Figure~\ref{fig:optspec}, top panel of Figure~\ref{fig:irspec}, and Figure~\ref{fig:esi}) 
shows lines like \ion{Ca}{1}, \ion{Mg}{1}, and \ion{Fe}{1} 
that are typically seen in cool photospheres.
However, the lines are shallow (few percent depths) and broadened in a non-gaussian way, 
more like line-splitting, which can be seen even in the only moderate spectral dispersion 
of our ESI spectrum.  Furthermore, there are many intermediate excitation lines
from \ion{Ni}{1}, \ion{Fe}{1}, and \ion{Fe}{2} (e.g. 5316, 6516 \AA), 
and clear indications of an even hotter spectral contributions as well. 
Absorption is obvious in many lines with excitation potential (EP) of 5-10 eV, 
for example \ion{Mg}{2} (e.g. 4481, 7877, 7896 \AA), 
\ion{Si}{1} (e.g. 6518, 8752 \AA) with 5.0 eV, the \ion{Ca}{2} 8912, 8927 \AA\ doublet 
\footnote{In the Sun, where they have a weak ($\sim 100$ m\AA) appearance, 
as well as most accretion-enhanced stars that display these lines, 
the 8927:8912 ratio is near unity but slightly weighted towards 8927.  
In \rno, however, it is actually the 8912 line that is much stronger than the 8927 line. 
\cite{hamann1992_aebe} noted the ``theoretical (optically thin) ratio of CaII 8927:8912 $\approx 1.4$ (Kurucz \& Peytremann 1975)," 
which suggests the lines are somewhat optically thick in \rno.}
as well as 9890 \AA\ (7.0 eV) 
and \ion{C}{1} lines in the 9061-9112 \AA\ range (7.5 eV). 
At the same time, there is a cool contribution as evidenced by weak TiO bandhead absorption at 8859 and 9209 \AA. 
Neither the hotter atomic lines nor the cooler molecules are seen in typical T Tauri stars,
such as BP Tau. While Figures~\ref{fig:optspec} and ~\ref{fig:esi} demonstrate many of these lines; not all are illustrated.

The infrared spectrum of \rno\ (Figure~\ref{fig:irspec})
shows clear molecular band absorption 
in TiO (Y band), CN (Y band), VO (Y and J bands), 
$H_2O$ (J, H, and K bands), and $CO$ (H and K bands). 
There are also atomic lines such as 
\ion{Na}{1}, \ion{Ca}{1}, \ion{Al}{1}, \ion{Mg}{1} and \ion{Si}{1}.
Notable are the higher-excitation atomic lines from \ion{Ca}{2} and \ion{C}{1} (around 9000 \AA).

Consistent with other FU Ori stars displaying a mixed spectrum with both
cooler and hotter opacity contributors, there is also strong evidence for 
low gravity.  In the optical, 
this is indicated by enhanced \ion{Ba}{2} 6141 and 6496 \AA\ absorption,
lines with positive luminosity effect \citep{andrievsky1998}, 
as well as the strengths of several \ion{Ti}{1} lines and the ratio 
of \ion{Fe}{2} 5316 \AA\ to \ion{Fe}{1} 5328 \AA\ that are highlighted by \cite{bob_carvalho2023b} as gravity-sensitive.
In the infrared, important indicators of low surface gravity are the \ion{Sr}{2} lines
in Y-band \citep{sharon2010}, CN in J-band \citep{wallace2000}, and
the strong \ion{Si}{1} line in the H-band \citep{meyer1998}.

Finally, \rno\ clearly shows \ion{Li}{1} 6707 \AA\ (see Figure~\ref{fig:profs}), a line which indicates stellar youth
and is detected in most young stellar objects (excepting those with very high veiling due to rapid accretion).
\ion{Li}{1} 6707 \AA\ is ubiquitously present in the extreme accretor FU Ori stars, 
and usually has a disk-like broadening often with additional kinematic signatures, perhaps from outflow.

\subsubsection{Wind/Outflow Lines}

As shown in Figure~\ref{fig:profs},
\rno\ exhibits H$\alpha$ emission with a P Cygni type profile, indicative of wind.  
The 10\% width of the emission part of the H$\alpha$ line is 170 km/s.
Blueshifted absorption dips are seen at about -90 and -175 km/s,
which also appear at a more muted level in the \ion{Ca}{2} triplet line profiles. 
The ``double-dip" structure on the blue side is not atypical for FU Ori type sources.
The higher Balmer series lines are in absorption, with asymmetric profiles
having extra blueshifted absorption.  

Wind signature is also seen in the absorption profiles of certain 
metal lines, including the deep \ion{Na}{1} D doublet at 55\% of continuum, 
the \ion{K}{1} 7669,7699 \AA\ doublet, and the \ion{O}{1} 7773 \AA\ triplet.
Furthermore, there is weak evidence in \rno\ for hot wind lines, such as 
\ion{Si}{2} 6347,6371 \AA\ (8.1 eV) that are prominent 
in early-stage FU Ori outbursts (see e.g. \citealt{bob_carvalho2023b} on V960 Mon).
The \ion{O}{1} 8446 \AA\ (9.5 eV) line is also present, but also weaker than in other examples of FU Ori stars.

In the infrared spectrum, 
there is a strong pure-absorption blue-asymmetric profile 
in the \ion{He}{1} 10830 line, with absorption out to -650 km/s. 
Absorption is present in the Paschen series from Pa$\beta$ up through the optical lines,
and in Br$\gamma$, but there is no signature in upper level Brackett series lines. 
The \ion{H}{1} absorption line widths are approximately 150 km/s (HWHD), and thus broader than the disk lines.

Overall, the presence of outflowing hot gas in the \rno\ system is clear. 
The wind profiles are not as deep nor as broadly and asymmetrically blueshifted 
as they appear in some other FU Ori stars, which we suspect is due to the relatively old age of the outburst.

In terms of jet activity, our spectrum does not show low-excitation forbidden lines 
such as [\ion{O}{1}] or [\ion{S}{2}]
that are often observed in ``low-state" accreting young stellar objects, 
and indicative of shocked material.  While
not typically seen in the ``high-state accretor" FU Ori stars, 
such lines have been observed in V1057 Cyg and V960 Mon,
FU Ori outburst sources that have faded substantially.
However, \cite{magakian2023} did identify optical
forbidden-line emission at spatially offset positions from the \rno\ point source,
suggesting possible past ejections.


\section{Accretion Disk Model Matched to \rno\ Spectra and SED}

In this section, we demonstrate that in addition to displaying the spectral
signatures of an FU Ori star, \rno\ is well-fit by a standard
pure-accretion disk model. 

\begin{figure}[!b]
\includegraphics[width=0.48\linewidth]{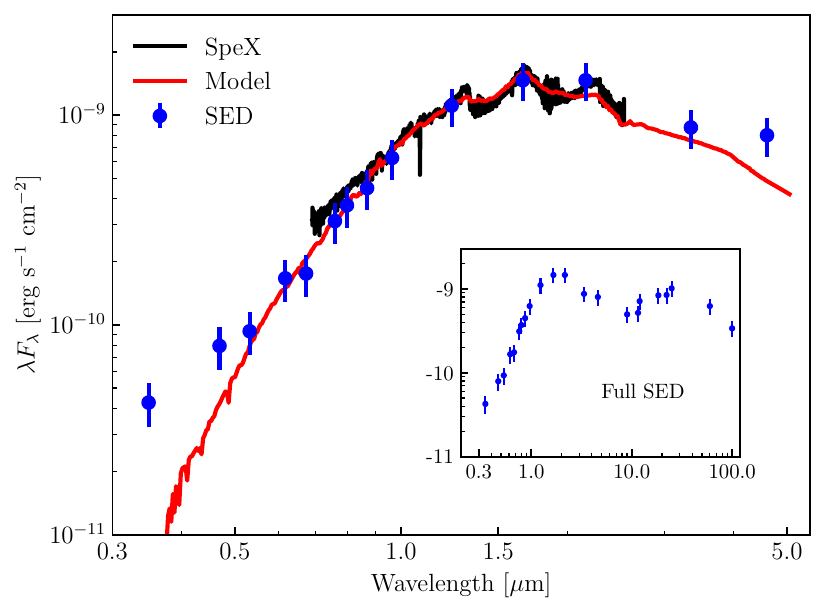}
\includegraphics[width=0.48\linewidth]{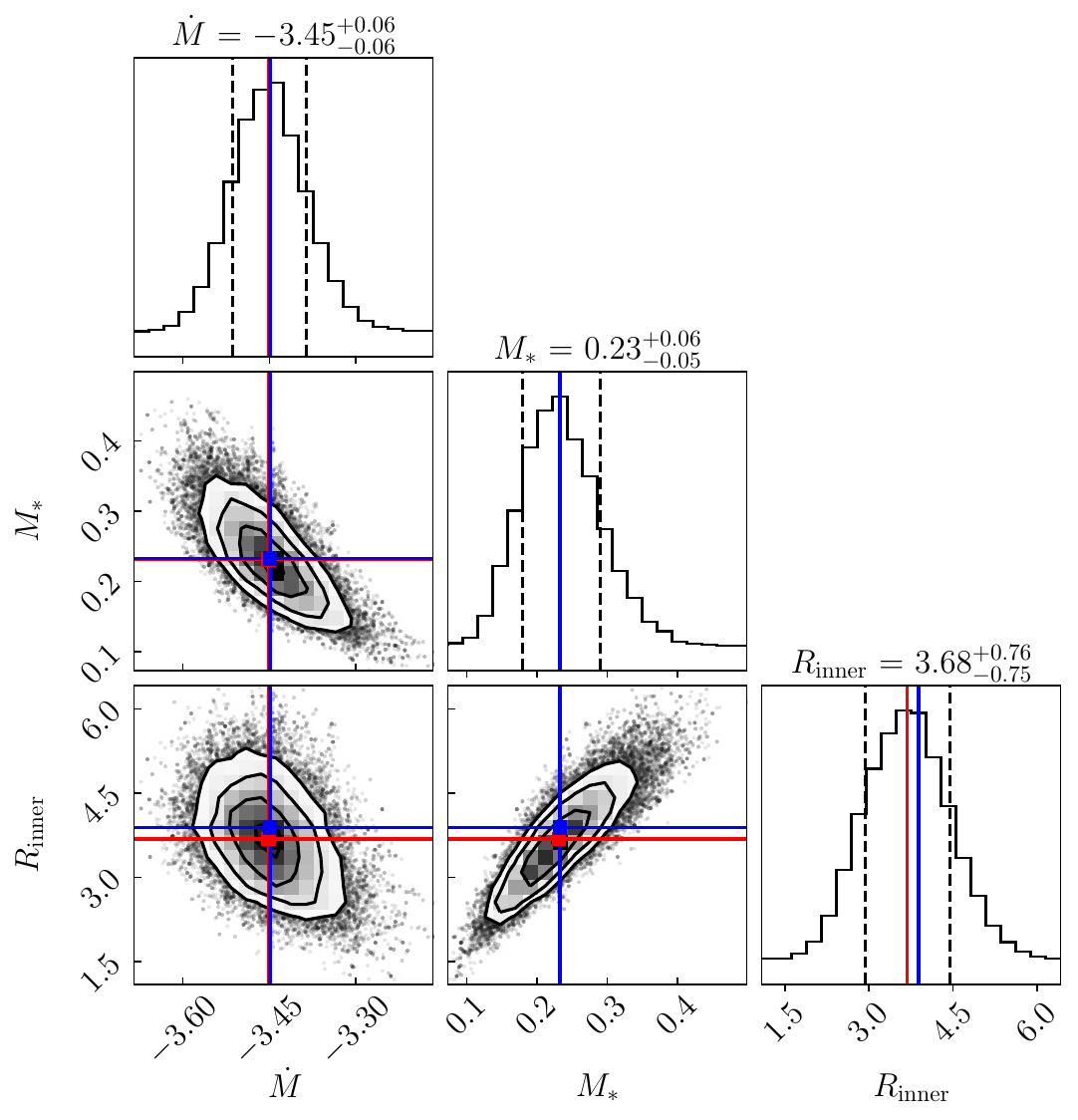}	
\caption{Left: assembled photometric SED (blue points), spectrophotometry from SpeX (black line), and accretion disk model fit (red line).  
Right: ``corner plot" showing the physical parameters in the model;
red lines mark median values while blue lines are modal values.
 The gas disk model is appropriate over the optical and near-infrared wavelength range,
where accretion dominates the emission. 
Beginning in the mid-infrared, dust contributions from an inactive or passive disk 
become important, while in the far-infrared dust emission from an envelope dominates;
these cooler contributions to the SED are not part of our modelling effort.
} 
\label{fig:sed}
\end{figure}

We assembled photometry and created the SED shown in Figure~\ref{fig:sed}.
The optical measurements are from UVEX ($u,g$) , IPHaS ($r,i$), PanSTARRS ($z,y$), 
and Gaia ($G,B_P,R_P$), while near-infrared measurements are from 2MASS ($J,H,K$). 
Mid-infrared data (shown in the inset) are from WISE, AKARI, and IRAS
indicate \rno\ has a Class I type spectral energy distribution, 
with spectral index +0.24 (rising to the red).
Catalog references for the above are \cite{mongui2020,drew2014,flewelling2020,gaiadr3,cutri2003,wise_data,akari,neugebauer1984}.

We fit the SED following a procedure similar to that described in \cite{carvalho2023a}. 
The priors we impose on the MCMC fit are: a Kroupa IMF constraining the central star mass, and a requirement imposed by the line broadening in the ESI spectrum, 
that $v_\mathrm{max} \times \sin i$ should be drawn from a Gaussian distribution centered on 85 km s$^{-1}$ with a standard deviation of 10 km s$^{-1}$. 
The $v_\mathrm{max}$ is calculated at each iteration as the Keplerian velocity of the innermost annulus of the disk: $v_\mathrm{max} = \sqrt{G M_*/R_\mathrm{inner}}$. 
Our fitting process initially considered the parameters:
disk accretion rate $\dot{M}$, stellar mass $M_*$, stellar radius $R_*$, disk inclination $i$, extinction $A_V$, and distance $d$.
The fit converged with decently well-behaved posteriors, but in order to better constrain the most important disk parameters,
we chose to adopt fixed values for $d$, $A_V$, and $i$.  
The initial distance posterior was $1490 \pm 670$ pc and we thus adopted $d = 1400$ pc, the Gaia DR3 value.
The initial extinction posterior was $3.93 \pm 0.67$ mag and we thus adopted 3.93 mag, which is consistent within the errors of
the value of 4.2 mag resulting from the DIBs measurement as well as Gaia DR3.
The initial inclination posterior was $47 \pm 14$ deg and we thus adopted 50 deg, also 
consistent with the morphology of optical and infrared images of the source that show a cometary morphology in its scattered light nebula.
In particular, it is the inclination constraint that helps narrow the parameters of the accretion disk.  

The posteriors for the remaining parameters are shown in the $\mathtt{corner}$ plot in Figure~\ref{fig:sed} for our final fit. 
The system parameters are: $\dot{M} = 10^{-3.45\pm0.06}$ $M_\odot$ yr$^{-1}$, $M_* = 0.23\pm0.06 \ M_\odot$, and $R_\mathrm{inner} = 3.68\pm0.76 \ R_\odot$. 
The resulting best-fit SED is a good match to the optical/NIR photometry and the SpeX spectrophotometry and is also shown in Figure~\ref{fig:sed}. 
 
We can derive additional parameters of the \rno\ system, namely the maximum temperature of the disk, $T_{max}$ and the accretion luminosity $L_{acc}$, 
via comparison with the V960 Mon system 
that was similarly modelled in detail by \cite{carvalho2023a}.
Using our infrared spectrum for \rno\ and that for V960 Mon from \cite{carvalho2023a} at the January 2016 epoch, 
when the object had cooled somewhat from the earlier outburst epoch,
we find that the objects closely match one another,  especially in many of the molecular features. 
We assume, therefore, that the $T_\mathrm{max}$ of the two objects should be approximately similar, 
and adopt for \rno\ the $T_\mathrm{max} \sim 7000$ K estimated in \cite{carvalho2023a} for V960 Mon at this late epoch. 
This temperature is somewhat cooler than the formal temperature following from the most likely system parameters given above,
but is within the $1\sigma$ values, notably reached by making $R_\mathrm{inner}$ larger. 

Continuing this line of argument,
we find that dereddening the RNO 54 spectrum by $A_V = 3.5$ mag gives a good match to the dereddened V960 Mon spectrum, which confirms our best-fit $A_V$ from the SED fitting. Assuming then a distance of 1.4 kpc to RNO 54 and an inclination of 50 deg, we find an integrated luminosity of $L_\mathrm{SpeX} \sim 137 \ L_\odot$. In V960 Mon, the $L_\mathrm{acc} \sim 2 \ L_\mathrm{SpeX}$, which would give 
$L_\mathrm{acc} \sim 265 \ L_\odot$ for RNO 54.
Fixing $T_\mathrm{max}$ and assuming $T_\mathrm{max} \propto \left({L_\mathrm{acc}}/{R_\mathrm{inner}^2}\right)^{1/4}$, we can then compute $R_\mathrm{inner} \sim 4.5 \ R_\odot$ and that $\dot{M}M_* \sim 7.72 \times 10^{-5} \ M_\odot^2$ yr$^{-1}$. These values are in good agreement with the system parameters we derived above from the SED fit. 
   

\section{Discussion}

\subsection{\rno\ in Context}

Compared to Gaia 18dvy and Gaia 17bpi, both recently outbursting FU Ori stars
that are also optically visible sources,
\rno\ seems to have broader and shallower absorption lines. 
This is probably due to its higher inclination, 
which we found in our disk modelling to be close to 50 deg.  
The line ratios, in particular the relative strengths of 
\ion{Fe}{1} and \ion{Fe}{2} lines, also suggests that \rno\ is hotter than
both Gaia 18dvy and Gaia 17bpi. For the same reasons, 
\rno\ is potentially cooler than V960 Mon, even at its later epochs, 
as evidenced by its stronger TiO and VO bands, as well as its
stronger \ion{Fe}{1}, and relatively weaker \ion{Fe}{2} lines.

V960 Mon is a recent outburst erupting in late 2014, but one that has faded and cooled quickly. 
\cite{bob_carvalho2023b} demonstrated the rapid evolution of its wind lines with
\ion{Si}{2} 6347,6371 \AA\ and \ion{O}{1} 8446 \AA\ weakening by a factor of three on a few year timescale
as V960 Mon cooled from the peak of its outburst.  In \rno, which 
is a much older FU Ori ouburst than V960 Mon, and observed spectroscopically only many decades 
after its assumed outburst, these lines along with \ion{H}{1} Paschen and Bracket absorption
are still present, demonstrating wind, but very weak. 
An exception is \ion{He}{1} 10830 \AA\ which is exceptionally strong in \rno.


In terms of other information on \rno, there is low level variability in the
optical lightcurves that are publically available from e.g. ASAS-SN, ZTF
and ATLAS.  The variations are quasiperiodic on long timescales (months)
with amplitudes of about 0.2 mag, which is nothing unusual for a YSO.
The mid-infrared lightcurve and colorcurve from NEOWISE are both relatively flat,
and again unremarkable.

\subsection{Comparison to \cite{magakian2023}}

Several parameters of \rno\ are found to be similar between the present study 
and that of \cite{magakian2023}.
The extinction was reported as $A_V = 2.5-3.0$ mag in \cite{magakian2023} 
whereas we find a value closer to 4 mag.
The luminosity was $L = 300-400 ~L_\odot$ in \cite{magakian2023}, 
whereas we find a lower $265 ~L_\odot$ for the inner accretion disk, 
though closer to 580 $L_\odot$ in the total SED including the mid- and far-infrared.

\subsection{Other Remarks}


It is not unreasonable to assume that \rno\ is only one of at least a few tens
of recognized YSOs that is actually an object experiencing an accretion outburst.
YSO selection techniques based on infrared excess as a disk indicator,
or activity diagnostitics such as x-ray emission, $H\alpha$ emission, or photometric variability, 
have identified several hundred thousand YSOs within 1-2 kpc of the Sun.  
However, a relatively small fraction of these have been studied 
spectroscopically.  

The tell-tale FU Ori signature at low-dispersion 
is a wavelength-dependent spectral type from optical FGK to near-infrared (M), 
and at high-dispersion, a mixed-temperature low-gravity spectrum over short
spectral ranges.  In spectroscopic studies that consider only optical data, or
only near-infrared data (but not both), this could render {\it bona fide} 
FU Ori stars removed from samples due to similarity to background giant contaminants
(FGK types in optical spectra, and M types in infrared spectra).

Given the recent discovery rate of approximately one new FU Ori outburst every 1-2 years,
and the lower rates in the past (one discovery only every $\sim$10 years),
there are likely to be a few tens of unrecognized ``FUOr-like" stars
in which no outburst was detected/noticed, but the object currently does
exhibit the spectral signatures of being in an outburst state.

\section{Summary}

In this paper, we have presented evidence for the young stellar object \rno\ 
as satisfying all of the commonly accepted criteria for a post-outburst FU Ori 
type star.  No actual photometric outburst was detected in this source, 
but the hypothesized burst is constrained to have occurred 
prior to the 1950s, and likely prior to the 1890s.
Evidence in favor of FU Ori status --
beyond the common YSO features of having a Class I SED showing 
broad infrared excess, a nebular environment, 
and the \ion{Li}{1} 6707 \AA\ spectral signatures of stellar youth -- includes:
\begin{itemize}
\item
An optical absorption spectrum featuring metal line absorption from a range of
hotter 
and cooler 
lines that are broadened to 85 km/s, consistent with the rotational broadening of a disk;
\item
An infrared spectrum featuring a similar mix of higher and lower excitation potential atomic lines,
as well as prominent molecular bands.
\item
Low surface-gravity indicators such as atomic \ion{Sr}{2} and \ion{Si}{1} lines, 
as well as molecular bands due to CN,
H$_2$O absorption sufficiently deep to produce a ``triangular H-band" spectral shape,
and strong CO absorption.
\item
Wind/outflow evident in the P Cygni type profile of H$\alpha$, 
and a hint of such in \ion{Ca}{2} triplet lines,
as well as blueshifted asymmetry in other atomic absorption line profiles, 
prominently \ion{He}{1} 10830 \AA. 
\item
The match of observed spectrophotometry to an accretion disk model with
system parameters: $\dot{M} = 10^{-3.45\pm0.06}$ $M_\odot$ yr$^{-1}$, $M_* = 0.23\pm0.06 \ M_\odot$, and $R_\mathrm{inner} = 3.68\pm0.76 \ R_\odot$;
we note, however, that $R_\mathrm{inner}$ is likely to be close to its upper range of $4.5 R_\odot$. 
The corresponding radiative parameters of the accretion disk are thus
$T_{max} \approx 7000$ K and $L_{acc} \approx 265 L_\odot$.
\end{itemize}

We speculate that \rno\ may be only one of at least a few tens
of known YSOs that is an object in an unrecognized state of accretion outburst.

\begin{acknowledgments}
We thank our referee for detailed comments that helped improve the manuscript, and 
Tigran Magakian for reminding us of the distinction between cometary and conical nebulae when describing YSOs.
\end{acknowledgments}

\vspace{5mm}
\facilities{Keck:II(ESI), IRTF(SpeX)}

\bibliography{ms}{}
\bibliographystyle{aasjournal}

\end{document}